# Case Study: Evaluation of a meta-analysis of the association between soy protein and cardiovascular disease

S. Stanley Young,[1] Warren B. Kindzierski,[2] Douglas Hawkins,[3] Paul Fogel,[4] Terry Meyer[5]

[1]CGStat, Raleigh, NC 27607, USA; genetree@bellsouth.net
[2]Independent Consultant, St Albert, Alberta, Canada T8N5R1; wbk@shaw.ca
[3]Emeritus Professor of Statistics, University of Minnesota, Scottsdale AZ 85262 USA; dhawkins@umn.edu
[4]Independent Consultant, Paris 75006, France; paul_fogel@hotmail.com
[5]Outcome Based Medicine, Raleigh, NC 27614, USA; terry_g_meyer@yahoo.com

## Abstract (190 words)

It is well-known that claims coming from observational studies most often fail to replicate. Experimental (randomized) trials, where conditions are under researcher control, have a high reputation and meta-analysis of experimental trials are considered the best possible evidence. Given the irreproducibility crisis, experiments lately are starting to be questioned. There is a need to know the reliability of claims coming from randomized trials. A case study is presented here independently examining a published meta-analysis of randomized trials claiming that soy protein intake improves cardiovascular health. Counting and p-value plotting techniques (standard p-value plot, p-value expectation plot and volcano plot) are used. Counting (search space) analysis indicates that reported p-values from the meta-analysis could be biased low due to multiple testing and multiple modeling. Plotting techniques used to visualize behavior of the data set used for meta-analysis suggest that statistics drawn from the base papers do not satisfy key assumptions of a random effects meta-analysis. These assumptions include using unbiased statistics all drawn from the same population. Also, publication bias is unaddressed in the meta-analysis. The claim that soy protein intake should improve cardiovascular health is not supported by our analysis.

# Introduction

***Irreproducible science***: It is known that research findings coming from observational studies, experimental studies (randomized trials) and, in general, other studies across scientific disciplines often fail to replicate. Some examples explain this point. Firstly, a 1996 example relates to *in vitro* activity testing of weak environmental estrogens – see Figure 1 (Arnold et al. 1996). Researchers claimed that pairs of essentially inactive estrogens – e.g., weak environmental estrogens such as dieldrin, endosulfan or toxaphene – were strongly estrogenic and could be potentially harmful to biological systems when combined. Six other labs subsequently attempted to test their claims, and none could replicate them (Kaiser 1997).

Secondly, Young and Karr (2011) observed that 0 of 52 medical treatment claims suggested by published observational studies failed to replicate when retested in randomized clinical trials (RCTs). Some of these claims came with superior credentials – small p-values, 0.001, published in prestigious journals such as the *New England Journal of Medicine* (e.g., Hu et al. 1997). This and other claims failed in large, well-designed randomized clinical trials, RCTs.

Thirdly, Begley and Ellis (2012) elaborated on the replication problem in the experimental biology field. They reported that the biotechnology firm Amgen (Thousand Oaks, CA) tried to confirm findings in 53 'landmark' studies published in the best biology journals. Their scientific findings were only able to confirm 6 of the 53 studies (11%).

Today there is acknowledged agreement that a replication problem exists in research across scientific disciplines (Chambers 2015, Hubbard 2015, Atmanspacher and Maasen 2016, NASEM 2016 and 2019, Harris 2017, Mayo 2018, Randall and Welser 2018, Ritchie 2020). Even in situations where a finding does replicate, it does not guarantee that such a finding is reliable (NASEM 2019). Negative studies are often not reported by authors (Franco et al. 2014). Even if negative studies are submitted for publication, editors may reject them out of hand, so a false positive finding can mistakenly be presumed as established fact (Franco et al. 2014, Nissen et al. 2016).

An often-cited cause of the replication problem has been statistical methods used by researchers (Colling and Szucs 2018). Nuzzo (2014) specifically identified significance testing and their associated p-values, and claims of statistical significance, as a key issue. Statistical inference plays an oversized role in science replicability due to frequent misuse of statistics – particularly the p-value and its threshold for determining statistical significance (NASEM 2019).

Small p-values do not provide a direct estimate of how likely a result is true or of how likely evidence against the null hypothesis ('there is an effect') is true. Further, small p-values do not convey whether a result is clinically or biologically important, large enough to have practical value. P-values are dependent on the data set, the statistical method used, assumptions made and the relevance of the assumptions (Chavalarias et al. 2016). NASEM (2019) further identified that inappropriate reliance on statistical significance may lead to biases in research reporting and publication.

*Meta-analysis*: A meta-analysis offers a window into a research claim, for example, that a foodstuff is causal of a chronic disease. A meta-analysis of this type examines a claim by taking a summary statistic along with a measure of its reliability from multiple individual nutritional survey–chronic disease studies (base papers) found in the epidemiological literature. These statistics are combined to give what is supposed to be a more reliable estimate of an air quality effect. Two key assumptions of meta-analysis are that: i) statistics drawn from base papers into the analysis are an unbiased estimate of the effect of interest (Boos and Stefanski 2013), and ii) meta-analysis of multiple studies offers a pooled estimate with increased precision (Cleophas and Zwinderman 2015) – i.e., there is one etiology in operation.

However, others have reported that the epidemiological literature is unbalanced (Thomas 1985, Ioannidis 2008, NASEM 2019). Negative studies – those with non-statistically significant results – are more likely to remain unpublished than studies with statistically significant results (Franco et al. 2014). Any systematic review or meta-analysis of these studies would be biased (Egger et al. 2001, Sterne et al. 2001) because they are combining information and data from a misleading, selected body of evidence (Thomas 1985, Ioannidis 2008, NASEM 2019). DerSimonian and Laird (1986) noted this publication bias problem in their influential paper presenting a random effects model for combining data in meta-analysis of randomized control trials,RCTs.

*Objective of study*: In this case study we chose to examine a meta-analysis of randomized studies investigating the association between soy protein and cardiovascular disease after Blanco Mejia et al. (2019). These randomized studies are being used by the US Food and Drug Administration (FDA) to decide whether to revoke a soy protein−heart health claim (US FDA 2017, Blanco Mejia et al. 2019).

## Methods

Blanco Mejia et al. (2019) performed a meta-analysis of 46 RCTs. The Blanco Mejia researchers observed soy protein intake and lipid markers (low-density lipoprotein or LDL cholesterol, and other cholesterol markers) as surrogates of cardiovascular disease (CVD) risk reduction in men and women. Our independent analysis focused on one aspect of their meta-analysis – LDL cholesterol as a surrogate of cardiovascular disease risk reduction.

Blanco Mejia et al. (2019) claimed "*...soy protein lowers LDL cholesterol by a small but significant amount. Our data fit with the advice given to the general public internationally to increase plant protein intake.*" This claim supports the original 1999 health claim (US FDA 2017). We used the same statistical strategy, analysis search space counting and p-value plots, to assess the Blanco Mejia et al. (2019) soy protein intake−LDL cholesterol health claim that we have used elsewhere (Young 2017, Young and Kindzierski 2019, Kindzierski et al., 2021).

*Data set*: Blanco Mejia et al. (2019) researchers reviewed all 46 dietary RCTs being used by the FDA in full. They selected 43 trials providing 50 study risk reduction comparison statistics to use for their meta-analysis. Epidemiologists traditionally use confidence intervals instead of p-values from a hypothesis test to demonstrate or interpret statistical significance. Since researchers construct both confidence intervals and p-values from the same data, the one can be calculated from the other (Altman and Bland 2011a,b). We extracted data from their Table 2 and computed p-values. All data used in our study is provided in Table 1.

*Search space counting*: To help assess the reliability of a statistical analysis it is useful to estimate (count) the number of questions that are at issue within a paper or report. A sound analysis will adjust p-values for the number of questions at issue. If a researcher computes many p-values (i.e., performs many statistical tests) on the same data set and makes no adjustment the researcher can be said to be p-hacking – trying this and that in search of a small p-value.

Scientists generally use a straightforward statistical analysis strategy on the data they collect – e.g., what causes or factors are related to what outcomes (health effects). This strategy allows researchers to analyze large numbers of possible relationships. If a data set contains "C" causes and "O" outcomes, then scientists can investigate C x O possible relationships. They can also examine how an adjustment factor "A" (also called a covariate), such as parental age, income, education, or parity of child, can modify each of the C x O relationships.

A 5 to 20% sample from a population whose characteristics are known is considered acceptable for most research purposes as it provides an ability to generalize for the population (Creswell (2003). We accepted the Blanco Mejia et al. (2019) judgment that their screening procedures selected 43 base papers with sufficiently consistent characteristics for use in meta-analysis. Based upon this, we randomly selected 9 of the Blanco Mejia et al. 43 base papers (21%) for counting purposes.

We counted the number of questions considered in base papers that we randomly selected for review. For this we counted causes (C), outcomes (O), and adjustment factors (A); where the number of questions = C x O × $2^A$.

*P-value plot*: We developed a p-value plot to inspect the distribution of the set of p-values after Schweder and Spjøtvoll (1982). A p-value plot is a method for correcting Multiple Testing and Multiple Modeling (MTMM) by allowing visual inspection the distribution of the set of p-values. The p-value is a random variable derived from a distribution of the test statistic used to analyze data and to test a null hypothesis.

In a well-designed study, the p-value is distributed uniformly over the interval 0 to 1 regardless of sample size under the null hypothesis and a distribution of true null hypothesis points plotted against their ranks in a p-value plot should form a straight line (Schweder and Spjøtvoll 1982,

Hung et al. 1997, Bordewijk et al. 2020). Researchers can use the plot to assess the reliability of base papers used in meta-analysis.

We constructed and interpreted p-value plots as follows (after Schweder and Spjøtvoll 1982, Young 2017, Young and Kindzierski 2019, Kindzierski et al., 2021):
- We computed and ordered p-values from smallest to largest and plotted them against the integers, 1, 2, 3, …
- If the points on the plot followed an approximate 45-degree line, we concluded that the p-values resulted from a random (chance) process, and that the data therefore supported the null hypothesis of no significant association.
- If the points on the plot followed approximately a line with slope < 1, where most of the p-values were small (less than 0.05), then the p-values provided evidence for a real (statistically significant) association.
- If the points on the plot exhibited a bilinear shape (divided into two lines), then the p-values used for meta-analysis are consistent with a two-component mixture and a general (overall) claim is not supported; in addition, the p-value reported for the overall claim in the meta-analysis paper cannot be taken as valid.
- We note that if the base papers are essentially perfect (randomization, no multiplicity problems, no selection bias, no fraud, etc.) and there is a real effect, then a bilinear p-value plot can result.

P-value plotting is not itself a cure-all. P-value plotting cannot detect every form of possible systematic error. Questionable research procedures and publication bias may alter a p-value plot (Young et al. 2021). But this plotting procedure it is a useful tool allowing one to detect a strong likelihood that questionable research procedures – such as data reliability, p-hacking, HARKing – may have distorted base studies used in meta-analysis and rendered meta-analysis unreliable.

P-hacking involves the persistent searching for statistical significance and comes in many forms, including multiple testing and multiple modeling, MTMM, without statistical correction (Ellenberg 2014, Hubbard 2015, Chambers 2017, Harris 2017, Streiner 2018). It enables researchers to find nominally statistically significant results even when there is no real effect (Boffetta et al. 2008, Ioannidis et al. 2011, McLaughlin and Tarone 2013, Simonsohn et al. 2014). To HARK is to *hypothesize after the results are known* – to look at the data first and then come up with a hypothesis, narrative, that has a statistically significant result (Randall and Welser 2018, Ritchie 2020).

P-value plotting is not the only means available by which to detect questionable research procedures. Other independent statistical tests may be available to account for frailties in base studies as they compute meta-analyses. Unfortunately, questionable research procedures in base studies severely degrade the utility of the existing means of detection (Carter 2019). We offer p-value plotting as one possible approach to detect questionable research procedures in meta-analysis.

***P-value expectation plot***: In a p-value expectation plot, the negative $\log_{10}$ of a p-value is plotted against its expected value assuming no effect. A p-value expectation plot allows a more careful examination of the distribution of the small p-values. The plot reverses the order of the p-values – small p-values appear in the upper right of the figure; those p-values that are close to their theoretical expectations, null p-values, appear near the dashed line.

***Volcano plot***: A volcano plot is a type of scatterplot used to examine patterns in data, particularly for identifying statistically significant differences between two populations (Hur et al. 2018). The volcano plot we used was constructed by plotting the negative $\log_{10}$ of the p value on the y axis against the relative risk (RR) statistic. The plot allows a visual examination of the p-values on a $\log_{10}$ scale along with the size and direction of a possible effect.

A consistent data set in a volcano plot should resemble an erupting volcano. Data points with low p-values (highly statistically significant results) appear toward the top of the plot. Data points in the top-right and top-left area of a volcano plot are of interest because they are the most different between the two conditions of interest – i.e., results with RRs >> 1 versus RRs << 1.

## Results

***Counting***: Summary search space characteristics for the nine RCT base study papers randomly selected from the Blanco Mejia et al. (2019) 43 base papers is presented in Table 2. The median search space of these nine RCT base study papers was 24. This is a much smaller number than a typical median search space of 10,000 or more for observational studies in the environmental epidemiology field (Young 2017, Young and Kindzierski 2019, Kindzierski et al. 2021).

Although no study is likely on its own to prove causality, randomization in RCT design is intended to reduce bias and provide a more rigorous means than observational studies for examining cause-effect relationships between a risk factor/intervention and an outcome (Hariton and Locascio 2018). Randomization promotes balancing of participant characteristics (both observed and unobserved) between the study groups. This promotes attribution of any differences in outcome to the study risk factor/intervention. In theory, results of a meta-analysis of RCTs should be superior to those from a meta-analysis based on observational studies.

***P-value plot***: The p-value plot for meta-analysis of the association between soy protein intake and LDL cholesterol reduction from Blanco Mejia et al. (2019) is shown in Figure 2. The p-value plot is clearly bilinear and hence ambiguous. Most of the p-values plot on a roughly 45-degree line.

***P-value expectation plot***: The expectation p-value plot for the Blanco Mejia et al. data is shown in Figure 3 with the small p-values appearing in the upper left of the figure. Those p-values close to their theoretical expectations, null p-values, appear near the dashed line. The dashed line is a linear fit to the values 1/(n+1), 2/(n+1), … n/(n+1), the expected values of the order statistics for the uniform distribution U(0, 1). The expected value of the smallest p-value from a U(0, 1) is 1/(n+1) where n = 50 or ~0.02, and we plot a dashed red line at $-\log_{10}$ of this value in Figure 3.

*Volcano plot*: The volcano plot for the Blanco Mejia et al. data is presented in Figure 4. A more typical volcano plot for a consistent set of data (i.e., drawn from the same population) should be a V−shaped distribution of points much like the points that appear in the lower right area of Figure 4. Several unusual patterns appear in Figure 4 indicating anomalies in the Blanco Mejia et al. (2019) data set:
- there are two very small p-values, having a $-\log_{10}$ between 6 and 7; these raw p-values are about $10^{-7}$; these p-values are from two studies that differ greatly in the RRs
- there is a cluster of four studies with $-\log_{10}$ p-values between 4 and 5 and RRs about 0.93

## Discussion

In addition to technical analysis aided by p-value plots, a meta-analysis of RCTs requires further careful judgement and thinking. In theory, the required conditions/assumptions of DerSimonian and Laird (1986) are much more likely to be met in aan RCT-based meta-analysis due to randomization and possible restrictions on the number of questions at issue. There should be unbiased treatment effects under those conditions.

On the other side of the ledger there can be MTMM problems with some of the studies and, of course, there can be publication bias. A first examination of the p-value plot (Figure 2) gives a bilinear plot. Under ideal conditions – DerSimonian and Laird (1986) assumptions satisfied, no publication bias and with a with a real (true) effect – a p-value plot will be bilinear, simulations, results not shown. Therefore, a discussion of assumptions underpinning the DerSimonian and Laird approach is warranted.

The DerSimonian and Laird (1986) approach is intended to address how results from small RCTs might be combined. That is, they assumed the numbers coming from individual RCTs are unbiased and that these individual RCTs were all measuring the same underlying process/effect (i.e., the numbers are all drawn from the same population). They also noted that their method does not take publication bias, no non-reporting of negative results, into account.

Non-significant studies are rarely reported (Franco et al. 2014). Greenwald (1975) puts the ratio of 'not reported' negative studies−to−reported positive studies at 10−to−1. In our situation, the Figure 4 volcano plot appears to be missing many points with p-values greater than 0.05 on the right side. Remember, studies with small p-values (less than 0.05) are, in theory, considered interesting and are more likely to be published and show up in a volcano plot; whereas studies with p-values greater than 0.05 are less likely to be published and not show up in the plot.

P-values less than 0.05 (i.e., studies that imply 'interesting' findings) warrant further discussion. The points on the blade of the hockey stick in the Figure 2 p-value plot – all p-values less than 0.05 – could be:
- valid, indicating a real (true) effect
- random, false-positive results
- moved to the blade through the practice of p-hacking, an essentially widespread practice (Head et al. 2015)
- the result of "data gardening" or even data fraud (e.g., see the retracted *Lancet* 2020 study of hydroxychloroquine (HQC) for treatment of COVID-19 (Mehra et al. 2020, Open Letter 2020))

Enstrom (2017) noted a historically important publication indicted that poor air quality causes increased mortality, and data gardening removed a portion of the data so that a 'not significant' finding became 'significant'. Enstrom (2017) obtained a copy of data that for many years was sequestered.

The two smallest p-values ($2.7 \times 10^{-7}$ and $3.3 \times 10^{-7}$) in Table 1 were small enough, well less than 0.001, that under perfect experimental conditions, would often be taken as indicating a real effect (Boos and Stefanski 2011). Here we must enter a note of caution as one may be viewing failures (violations) of research integrity (Roberts et al. 2007, Redman 2013, Bordewijk et al. 2020).

Tugwell and Knottnerus (2017) classify research integrity violations as:
- data fabrication (e.g., use of data from an uncredited author or generation of completely artificial data)
- data falsification (e.g., editing or manipulation of authentic data to ''support'' a hypothesis)
- unethical author conduct (e.g., failure to obtain institutional review board approval, failure to obtain patient informed consent, forgery of secondary authors' signatures on submission, other breaches of ethical guidelines)
- error (e.g., duplicate publication, scientific mistake, journal error, unstated reasons for retraction)

Roberts et al. (2007), Redman (2013) and Bordewijk et al. (2020) contend that integrity violations in research are substantially more common than what many researchers think. They contend that this is facilitated by systematically lax oversight by journals and institutions – not least their failure to require researchers to provide their data sets.

In our case study, there are seven p-values that are rather clearly not like the rest. Characteristics of these seven p-values are presented in Table 3. These seven p-values are both outliers and influential in a meta-analysis. The points in the lower right area are of Figure 4 give the more typical appearance of a volcano plot – as the points diverge from RR of 1.0, they go up to the left and up to the right. With the seven outlier values removed – refer to Figure 5 – the pattern gives an impression of no overall effect. Again, it is acknowledged that Figure 5 may not even be representative, as points indicative of 'not reported' negative studies that should show up on the right on the right in Figure 5 are likely unreported.

## Findings

Counting (search space) analysis indicates that reported p-values from Blanco Mejia et al. could be biased low due to multiple testing and multiple modeling. P-value plotting techniques used to visualize behavior of the Blanco Mejia et al. data set suggest that statistics drawn from the base papers do not satisfy key assumptions of a random effects meta-analysis. These assumptions include using unbiased statistics all drawn from the same population. Also, publication bias is unaddressed in the meta-analysis. The claim that soy protein intake should improve cardiovascular health is not supported by our analysis.

**Figures**

Figure 1. Irreproducible endocrine disruption research.

> In 1996 several claims were advanced by Tulane University researchers in the journal *Science* that pairs of essentially inactive estrogens – e.g., weak environmental estrogens such as dieldrin, endosulfan or toxaphene – were strongly estrogenic and could be potentially harmful to biological systems when combined (Arnold et al. 1996). Their claims were startling and potentially explanatory for all manner of biological harms in the environment.
>
> Six labs subsequently attempted to test their claims, and none could replicate them (Kaiser 1997). The Tulane researchers even tried to replicate their own claims but could not. They offered a possible explanation that there must have been a fundamental flaw in the design of their original experiment. They formally withdrew their paper in 1997 (McLachlan 1997). Funders of their original research – the US Public Health Service (PHS) – determined scientific misconduct by the Tulane researchers and barred from them receiving PHS grants for five years (Malakoff et al. 2001).
>
> Now consider a possible statistical explanation for their irreproducible findings… The Tulane researchers looked at many pairs of compounds. For example, 20 compounds will generate 190 possible pairs, 40 compounds will generate 780 possible pairs. With many possible pairs to examine, their 1996 *Science* paper was not explicit on the actual number of pairs that they examined. Chance could produce extreme results and attempts to replicate results for only specific pairs would be expected to fail. It seems apparent that the Tulane researchers did not replicate their own findings before they submitted their paper to *Science*.
>
> Their original *Science* paper (Arnold et al. 1996) has been cited 768 times as of 4 November 2021. Their formal withdrawal notification (McLachlan 1997) has been cited 172 times. The whole idea of endocrine disruptors has since taken on something of a life of its own with the term 'endocrine disruptors' in the title of over 2,000 published papers as of 4 November 2021. The US Environmental Protection Agency has an ongoing program of endocrine disruptors (https://www.epa.gov/endocrine-disruption).

Figure 2. P-value plot for meta-analysis of the association between soy protein intake and LDL cholesterol reduction from Blanco Mejia et al. (2019).

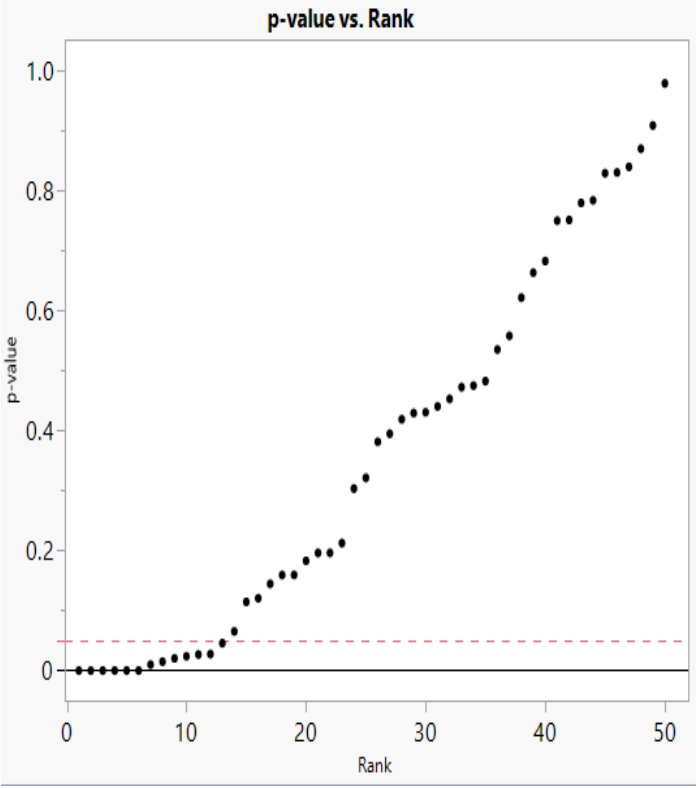

Figure 3. Expectation p-value plot for meta-analysis of the association between soy protein intake and LDL cholesterol reduction from Blanco Mejia et al. (2019).

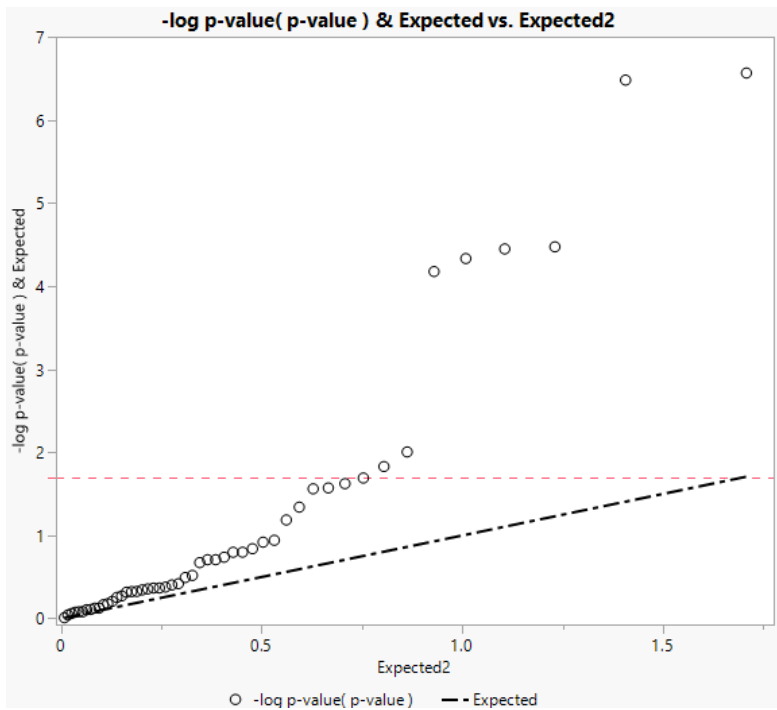

Note: The expected value of the smallest p-value from a uniform distribution U(0, 1) is $1/(n+1)$ where n = 50 or ~0.02, and a dashed red line is plotted at −log10 of this value.

Figure 4. Volcano p-value plot showing the negative log10 of the p-values plotted against the risk ratio.

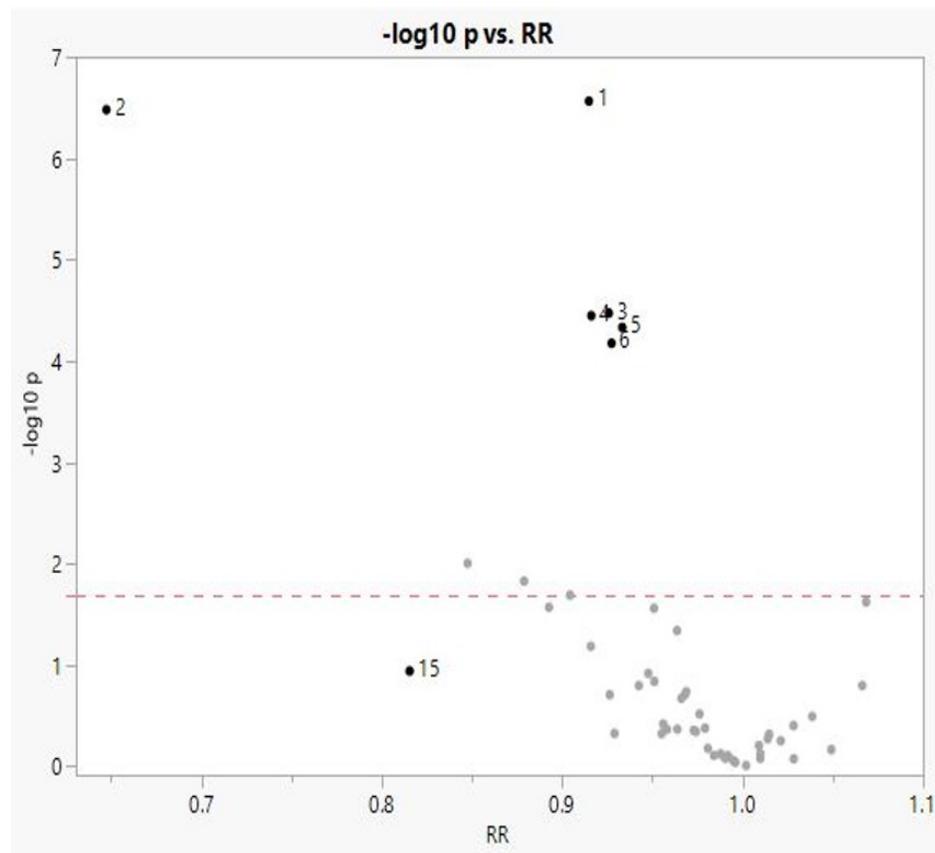

Note: The expected value of the smallest p-value from a uniform distribution U(0, 1) is 1/(n+1) where n = 50 or ~0.02, and a dashed red line is plotted at −log10 of this value.

Figure 5. Volcano p-value plot with the Table 3 seven outlier p-values removed showing the negative $\log_{10}$ of the remaining p-values plotted against the risk ratio.

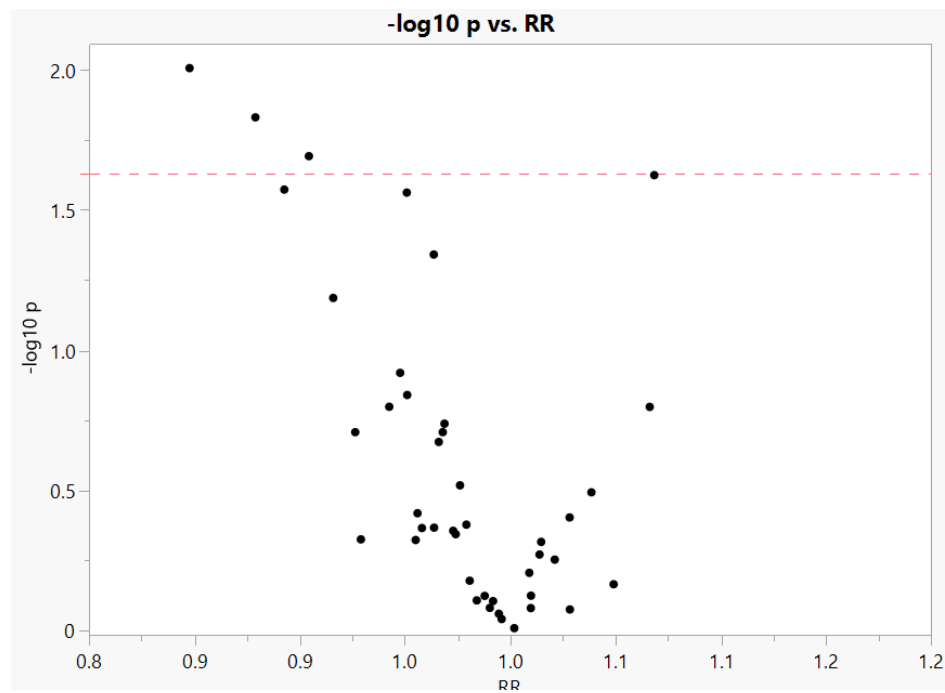

Note: The expected value of the smallest p-value from a uniform distribution U(0, 1) is 1/(n+1) where n = 43 or ~0.023, and a dashed red line is plotted at −log10 of this value.

# Tables

Table 1. Lists of studies used by Blanco Mejia et al. (2019) along with extracted and computed statistics.

| Author | Year | Comment | Ref | RR | CL$_{low}$ | CL$_{high}$ | SE | Z | p-value | Rank |
|---|---|---|---|---|---|---|---|---|---|---|
| Bakhit | 1994 | cotyledon | 15 | 1.0098 | 0.9497 | 1.0699 | 0.0307 | 0.3203 | 0.748745 | 41 |
| Bakhit | 1994 | cellulose | 15 | 0.9510 | 0.8852 | 1.0168 | 0.0336 | -1.4596 | 0.144396 | 17 |
| Blum | 2003 | | 16 | 1.0282 | 0.9634 | 1.0930 | 0.0331 | 0.8521 | 0.394158 | 27 |
| Bosello | 1988 | | 17 | 0.8926 | 0.7976 | 0.9877 | 0.0485 | -2.2147 | 0.026781 | 11 |
| Chen | 2006 | | 19 | 0.8476 | 0.7318 | 0.9634 | 0.0591 | -2.5796 | 0.009891 | 7 |
| Cuevas | 2003 | | 20 | 0.9959 | 0.9269 | 1.0649 | 0.0352 | -0.1167 | 0.907124 | 49 |
| Evans | 2007 | | 21 | 1.0211 | 0.9506 | 1.0915 | 0.0359 | 0.5873 | 0.55703 | 37 |
| Gardner | 2001 | | 23 | 1.0139 | 0.9701 | 1.0577 | 0.0223 | 0.6214 | 0.534318 | 36 |
| Gardner | 2007 | | 22 | 0.9508 | 0.9071 | 0.9945 | 0.0223 | -2.2050 | 0.027454 | 12 |
| Goldberg | 1982 | high chol | 24 | 0.9476 | 0.8816 | 1.0137 | 0.0337 | -1.5531 | 0.120402 | 16 |
| Goldberg | 1982 | normal chol | 24 | 1.0283 | 0.7564 | 1.3002 | 0.1387 | 0.2040 | 0.838334 | 47 |
| Greany | 2004 | | 26 | 0.9660 | 0.9126 | 1.0194 | 0.0272 | -1.2475 | 0.212208 | 23 |
| Harrison | 2004 | | 27 | 1.0662 | 0.9740 | 1.1585 | 0.0470 | 1.4078 | 0.159198 | 19 |
| Higashi | 2001 | | 28 | 0.9841 | 0.8731 | 1.0950 | 0.0566 | -0.2816 | 0.778243 | 43 |
| Høie | 2005 | 1 | 29 | 0.9148 | 0.8824 | 0.9473 | 0.0166 | -5.1416 | 2.72E-07 | 1 |
| Høie | 2005 | 2 | 30 | 0.9274 | 0.8916 | 0.9630 | 0.0182 | -3.9876 | 6.67E-05 | 6 |
| Høie | 2006 | active trt1 | 31 | 1.0147 | 0.9738 | 1.0555 | 0.0208 | 0.7036 | 0.481688 | 35 |
| Høie | 2006 | active trt2 | 31 | 0.9679 | 0.9193 | 1.0165 | 0.0248 | -1.2935 | 0.195823 | 21 |
| Høie | 2007 | | 32 | 0.9333 | 0.9012 | 0.9654 | 0.0164 | -4.0723 | 4.65E-05 | 5 |
| Hori | 2001 | | 33 | 0.6474 | 0.5120 | 0.7828 | 0.0691 | -5.1047 | 3.31E-07 | 2 |
| Jayagopal | 2002 | | 34 | 0.8789 | 0.7811 | 0.9760 | 0.0497 | -2.4370 | 0.01481 | 8 |
| Jenkins | 2000 | | 36 | 0.9580 | 0.8537 | 1.0622 | 0.0532 | -0.7893 | 0.429931 | 30 |
| Jenkins | 2002 | | 35 | 1.0385 | 0.9624 | 1.1146 | 0.0388 | 0.9928 | 0.32082 | 25 |
| Jenkins | 1989 | | 37 | 0.8154 | 0.5863 | 1.0445 | 0.1169 | -1.5792 | 0.114299 | 15 |
| Kohno | 2006 | 12 | 38 | 1.0684 | 1.0091 | 1.1277 | 0.0303 | 2.2604 | 0.023798 | 10 |
| Kohno | 2006 | 22 | 38 | 0.9946 | 0.9301 | 1.0590 | 0.0329 | -0.1657 | 0.868421 | 48 |
| Lichtenstein | 2002 | isoflavones | 39 | 0.9638 | 0.8741 | 1.0535 | 0.0458 | -0.7916 | 0.428569 | 29 |
| Lichtenstein | 2002 | no isoflavones | 39 | 0.9807 | 0.8941 | 1.0673 | 0.0442 | -0.4368 | 0.662248 | 39 |
| Liu | 2012 | | 40 | 0.9761 | 0.9307 | 1.0216 | 0.0232 | -1.0303 | 0.302854 | 24 |
| Ma | 2005 | | 41 | 1.0090 | 0.9733 | 1.0448 | 0.0182 | 0.4949 | 0.620668 | 38 |
| Maesta | 2007 | soy | 42 | 0.9043 | 0.8234 | 0.9852 | 0.0413 | -2.3195 | 0.020366 | 9 |
| Maesta | 2007 | soy+resis exer | 42 | 0.9550 | 0.8319 | 1.0782 | 0.0628 | -0.7157 | 0.474181 | 34 |

Note: Author is the 1st author listed for reference; Year = publication year; Comment describes multiple uses from the same paper; Ref# is Blanco Mejia et al. reference number; RR, CL$_{low}$, CL$_{high}$ are the risk ratios and their confidence limits; SE is standard error; Z is a z statistic of RR verses 1.00; p-value is for Z; Rank is the rank order of the p-values.

Table 1. Lists of studies used by Blanco Mejia et al. (2019) along with extracted and computed statistics (con't).

| Author | Year | Comment | Ref | RR | CL$_{low}$ | CL$_{high}$ | SE | Z | p-value | Rank |
|---|---|---|---|---|---|---|---|---|---|---|
| Mangano | 2013 | soy+isoflavone | 43 | 0.9729 | 0.9040 | 1.0416 | 0.0351 | -0.7728 | 0.439637 | 31 |
| Mangano | 2013 | soy+placebo | 43 | 0.9878 | 0.9130 | 1.0627 | 0.0382 | -0.3187 | 0.749918 | 42 |
| Matthan | 2007 | | 44 | 0.9791 | 0.9286 | 1.0296 | 0.0258 | -0.8096 | 0.418143 | 28 |
| McVeigh | 2006 | | 45 | 0.9161 | 0.8763 | 0.9558 | 0.0203 | -4.1339 | 3.57E-05 | 4 |
| Murkies | 1995 | | 47 | 1.0019 | 0.8723 | 1.1315 | 0.0661 | 0.0286 | 0.977221 | 50 |
| Murray | 2003 | | 48 | 0.9290 | 0.7357 | 1.1223 | 0.0986 | -0.7198 | 0.471677 | 33 |
| Santo | 2008 | | 50 | 1.0491 | 0.8146 | 1.2836 | 0.1196 | 0.4104 | 0.681537 | 40 |
| Steinberg | 2003 | | 51 | 0.9741 | 0.9064 | 1.0417 | 0.0345 | -0.7519 | 0.45214 | 32 |
| Takatsuka | 2000 | | 52 | 0.9159 | 0.8264 | 1.0053 | 0.0456 | -1.8441 | 0.065161 | 14 |
| Teede | 2001 | | 53 | 0.9636 | 0.9280 | 0.9993 | 0.0182 | -1.9988 | 0.045631 | 13 |
| Teixeira | 2004 | | 55 | 0.9687 | 0.9227 | 1.0147 | 0.0235 | -1.3325 | 0.182682 | 20 |
| Van Horn | 2001 | oats | 56 | 0.9903 | 0.9027 | 1.0779 | 0.0447 | -0.2177 | 0.827663 | 45 |
| Van Horn | 2001 | wheat | 56 | 1.0097 | 0.9214 | 1.0980 | 0.0450 | 0.2160 | 0.82899 | 46 |
| van Raaij | 1981 | | 57 | 0.9258 | 0.8908 | 0.9609 | 0.0179 | -4.1474 | 3.36E-05 | 3 |
| Washburn | 1999 | | 58 | 0.9425 | 0.8624 | 1.0225 | 0.0408 | -1.4083 | 0.159052 | 18 |
| West | 2005 | | 59 | 0.9918 | 0.9335 | 1.0501 | 0.0297 | -0.2760 | 0.782581 | 44 |
| Wong | 1998 | high chol | 60 | 0.9559 | 0.8573 | 1.0545 | 0.0503 | -0.8765 | 0.380764 | 26 |
| Wong | 1998 | normal chol | 60 | 0.9263 | 0.8146 | 1.0379 | 0.0570 | -1.2934 | 0.195875 | 22 |

Note: Author is the 1$^{st}$ author listed for reference; Year = publication year; Comment describes multiple uses from the same paper; Ref# is Blanco Mejia et al. reference number; RR, CL$_{low}$, CL$_{high}$ are the risk ratios and their confidence limits; SE is standard error; Z is a z statistic of RR verses 1.00; p-value is for Z; Rank is the rank order of the p-values.

Table 2. Results of counting outcomes, causes, covariates of nine randomly selected studies used by Blanco Mejia et al. (2019).

| Ref# | Author | Year | Outcomes | Causes | Adjustment factors (covariates) | Tests | Models | Search space |
|---|---|---|---|---|---|---|---|---|
| 15 | Bakhit | 1994 | 8 | 5 | 3 | 40 | 8 | 320 |
| 19 | Chen | 2006 | 9 | 1 | 4 | 9 | 16 | 144 |
| 33 | Hori | 2001 | 20 | 1 | 0 | 20 | 1 | 20 |
| 36 | Jenkins | 2000 | 14 | 1 | 0 | 14 | 1 | 14 |
| 41 | Ma | 2005 | 20 | 6 | 0 | 120 | 1 | 120 |
| 43 | Mangano | 2013 | 6 | 3 | 0 | 18 | 1 | 18 |
| 52 | Takatsuka | 2000 | 10 | 2 | 0 | 20 | 1 | 20 |
| 56 | Van Horn | 2001 | 3 | 4 | 1 | 12 | 2 | 24 |
| 60 | Wong | 1998 | 14 | 4 | 3 | 56 | 8 | 448 |

Note: Ref# is Blanco Mejia et al. reference number; Author name is first author listed for reference; Year = publication year; Tests = Outcomes × Causes; Models = $2^k$ where k = number of covariates; Search Space = approximation of analysis Search space = Tests × Models.

Table 3. Characteristics of seven outlier (influential) studies used by Blanco Mejia et al. (2019).

| Rank | Author | Year | Ref | RR | $CL_{low}$ | $CL_{high}$ | SE | Z | p-value |
|---|---|---|---|---|---|---|---|---|---|
| 1 | Høie | 2005 | 29 | 0.9148 | 0.8824 | 0.9473 | 0.0166 | -5.1416 | 2.72E-07 |
| 2 | Hori | 2001 | 33 | 0.6474 | 0.5120 | 0.7828 | 0.0691 | -5.1047 | 3.31E-07 |
| 3 | van Raaij | 1981 | 57 | 0.9258 | 0.8908 | 0.9609 | 0.0179 | -4.1474 | 3.36E-05 |
| 4 | McVeigh | 2006 | 45 | 0.9161 | 0.8763 | 0.9558 | 0.0203 | -4.1339 | 3.57E-05 |
| 5 | Høie | 2007 | 32 | 0.9333 | 0.9012 | 0.9654 | 0.0164 | -4.0723 | 4.65E-05 |
| 6 | Høie | 2005 | 30 | 0.9274 | 0.8916 | 0.9630 | 0.0182 | -3.9876 | 6.67E-05 |
| 7 | Jenkins | 1989 | 37 | 0.8154 | 0.5863 | 1.0445 | 0.1169 | -1.5792 | 0.114299 |

Note: Author is the 1st author listed for reference; Year = publication year; Ref# is Blanco Mejia et al. reference number; RR, $CL_{low}$, $CL_{high}$ are the risk ratios and their confidence limits; SE is standard error; Z is a z statistic of RR verses 1.00; p-value is for Z.